# A Simplified Multifractal Model for Self-Similar Traffic Flows in High-Speed Computer Networks Revisited

Ginno Millán[1], Gastón Lefranc[2]

[1] Universidad San Sebastián
[2] Pontificia Universidad Católica de Valparaíso

ginno.millan@uss.cl, gaston.lefranc@pucv.cl

**Abstract.** In the context of the simulations carried out using a simplified multifractal model that is proposed to give an explanation to the locality phenomenon that appears in the estimation of the Hurst exponent in the second-order stationary series that represent the self-similar traffic flows in high-speed computer networks, its formulation is perfected to reduce the variability in the singularity limits and it is demonstrated through by its wavelet variant that this modification leads to a higher resolution in the interval of interest under study.

# Revisión de un Modelo Multifractal Simplificado para Flujos de Tráfico Autosimilares en Redes de Computadoras de Alta Velocidad

**Resumen.** En el contexto de las simulaciones llevadas a cabo usando un modelo multifractal simplificado que se plantea para dar una explicación al fenómeno de localidad que aparece en la estimación del exponente de Hurst en series estacionarias de segundo orden que representan los flujos de tráfico autosimilares en redes de computadoras de alta velocidad, se perfecciona su formulación para reducir la variabilidad en los entornos de la singularidad y se muestra a través de su variante wavelet que dicha modificación implica una resolución mayor en el intervalo de interés bajo estudio.

**Palabras clave.** Redes de computadoras, exponente de Hurst ($H$), fenómeno de localidad, multifractales, modelos de tráfico.

## 1 Introduction

The properties that evidence the fractal nature of traffic flows in high-speed computer networks have been widely studied and reported in the literature during the last twenty years, and it is generally accepted that their rescaled dynamic behavior must be carefully considered in performance analyses [1].

Thus, there are numerous models that attempt to give an answer to this origin, e.g. [1-4].

On the other hand, admitting that the localities of a fractal process can only be analyzed from the standpoint of multifractal analysis, in view of their construction from the multiplicative cascades that ensure an exact characterization as a result of the high frequency analysis [5, 6], it is accepted that the traffic flows present in high-speed computer networks are of a multifractal behavior, and this gives rise to a new simplified multifractal model for traffic flows originally reported in [1], which gives an explanation for the locality phenomenon in the estimation of the Hurst exponent ($H$) [7, 8].

From the results obtained in [1] through the use of computational simulations, it is inferred that the model contributes to knowledge of traffic dynamics in current high-speed computer networks and can be used to simulate the approximate behavior of the real traffic flows. However its application deals with the totality of the curve of variations of Hurst exponent, which, although it reflects the locality phenomenon, widens the interest interval, which should be restricted to the neighborhood of the point where the curves shown in Figure 1 change their slopes and therefore the singularity occurs.

Furthermore, in [1] using representative fGn series, it is proves, as shown in Figures 2 and 3, that the model can capture the main trends of flows in the estimation of $H$.

However, the fact remains that the range of interest is too wide to capture the singularity and its neighborhood.

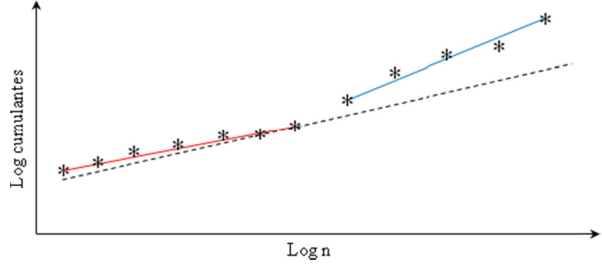

**Figure 1.** Illustration of the locality phenomenon.

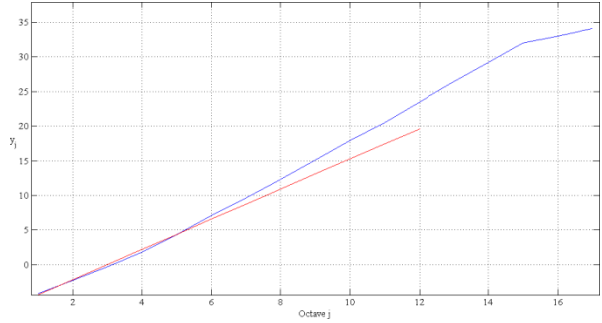

**Figure 2.** Locality phenomenon in a fGn ($H = 0.6$) [1]

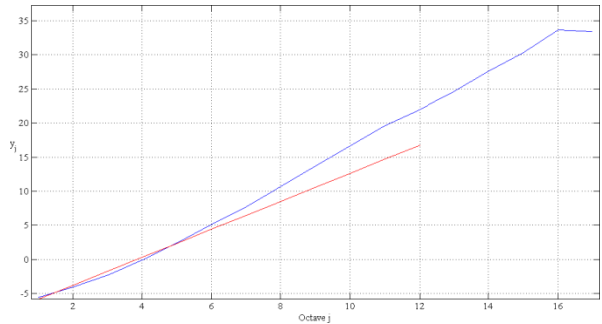

**Figure 3.** Locality phenomenon in a fGn ($H = 0.8$) [1]

## 2 Multifractal Traffic Flows Basics

In computer networks traffic flows is represented by means of a self-similar process $Y(k)$ such that

$$Y(k) =_d a^{-H} Y(ak), \quad \forall a > 0,\ k \geq 0, \quad (1)$$

where $=_d$ represents the equal finite-dimensional distributions and $H \in (0,1)$ is the Hurst exponent of the self-similar stochastic process ($H_{SS}$) $Y(k)$.

An $H_{SS}$ process with stationary increments in terms of the behavior of its aggregations is obtained by multiplexing of $X(k) = Y(k+1) - Y(k)$ increments over non-superposed blocks of size $n$ according to

$$X^{(n)}(k) = \sum_{j=0}^{n-1} X(kn - j), \quad k \in \mathbb{Z},\ n \in \mathbb{N}. \quad (2)$$

The resultant process has finite dimensional distributions similar to $X(k)$. Specifically for each $n$ we have that

$$X^{(n)}(k) =_d n^H X(k), \quad n \in \mathbb{N}. \quad (3)$$

The stationary process $X(k)$ that fulfills (3) is called a self-similar stationary process, $H$-SSS, with Hurst exponent $H$. A typical example is the fGn given by $X(k) = B^{(H)}(k+1) - B^{(H)}(k)$, which is the only known $H_{SSS}$ Gaussian process [9].

There are several ways to study the statistical properties of $X^{(n)}(k)$. However, the cumulants of the aggregate series are considered, which are defined as the Taylor coefficients of the cumulant function, as the best representation, given in [9] by

$$g(t) = \log E(e^{tX(k)}) = \sum_{m=1}^{\infty} t^m m!^{-1} \mathrm{cum}_m X(k), \quad (4)$$

where $\mathrm{cum}_m X(k) = g^{(m)}(0)$. In [10] it is shown that the $m$th cumulants of an $H_{SSS}$ aggregate process scales according to a power-law given by

$$\mathrm{cum}_m X^{(n)}(k) = n^{mH_S} \mathrm{cum}_m X(k). \quad (5)$$

If a $H_{SSS}$ process fulfills (5) $\forall\ n, m \in \mathbb{N}$, then $|\log \mathrm{cum}_m X^{(n)}(k)|$ behaves in such a way that its values scale linearly with those of $\log n$, with $mH_S$ coefficients that are a linear function of $m$. In other words, $mH_S = mH(m)$, i.e.

$$\mathrm{cum}_m X^{(n)}(k) = n^{mH(m)} \mathrm{cum}_m X(k). \quad (6)$$

In [5] it is shown that a generalization of a self-similar process to a multifractal process is given as follows: a stationary process $X(k)$ with $k \in \mathbb{Z}$ is a multifractal process if

$$\log |\mathrm{cum}_m X^{(n)}(k)| = mH(m)\log n + c(m), \quad (7)$$

for every $m, n \in \mathbb{N}$, allowing the exponent $H$ to vary with the order $m$.